\documentclass[10pt,preprint]{aastex}
\usepackage{epsfig}
\usepackage{amsmath}
\usepackage{multirow}
\usepackage{apjfonts}
\usepackage{color}

\begin{document}

\title{Constraints on the source of ultra-high energy cosmic rays
  using anisotropy vs chemical composition}

\author{Ruo-Yu Liu$^{1,2,3,*}$, Andrew M. Taylor$^4$, Martin
  Lemoine$^{5,\dagger}$, Xiang-Yu Wang$^{1,3}$ and Eli Waxman$^6$}
\affil{$^1$School of Astronomy and Space Science, Nanjing University, Nanjing, 210093, China\\
  $^2$Max-Planck-Institut f\"ur Kernphysik, 69117 Heidelberg, Germany\\
  $^3$Key laboratory of Modern Astronomy and Astrophysics (Nanjing University), Ministry of Education, Nanjing 210093, China\\
  $^4$Dublin Institute for Advanced Studies, 31 Fitzwilliam Place, Dublin 2, Ireland\\
  $^5$Institut d'Astrophysique de Paris, CNRS, UPMC, 98 bis boulevard Arago, F-75014 Paris, France\\
  $^6$Physics Faculty, Weizmann Institute, PO Box 26, Rehovot 7600, Israel\\
  $^*$Fellow of the International Max Planck Research School for Astronomy and Cosmic Physics at the University of Heidelberg (IMPRS-HD)\\
  $\dagger$ Email: lemoine@iap.fr}

\begin{abstract}
  The joint analysis of anisotropy signals and chemical composition of
  ultra-high energy cosmic rays offers strong potential for shedding
  light on the sources of these particles. Following up on an earlier
  idea, this paper studies the anisotropies produced by protons of
  energy $>E/Z$, assuming that anisotropies at energy $>E$ have been
  produced by nuclei of charge $Z$, which share the same magnetic
  rigidity. We calculate the number of secondary protons produced
  through photodisintegration of the primary heavy nuclei. Making the
  extreme assumption that the source does not inject any proton, we
  find that the source(s) responsible for anisotropies such as
  reported by the Pierre Auger Observatory should lie closer than
  $\sim$20-30, 80-100 and 180-200\,Mpc if the anisotropy signal is
  mainly composed of oxygen, silicon and iron nuclei respectively.  A
  violation of this constraint would otherwise result in the secondary
  protons forming a more significant anisotropy signal at lower
  energies. Even if the source were located closer than this distance,
  it would require an extraordinary metallicity $\gtrsim 120, 1600,
  1100$ times solar metallicity in the acceleration zone of the source, for
  oxygen, silicon and iron respectively, to ensure that the
  concomitantly injected protons not to produce a more significant low
  energy anisotropy. This offers interesting prospects for
  constraining the nature and the source of ultra-high energy cosmic
  rays with the increase in statistics expected from next generation 
  detectors.
\end{abstract}

\section{Introduction}
The origin of ultra--high energy cosmic rays (UHECRs) is a
long--standing puzzle of high energy astrophysics and astroparticle
physics. It is commonly believed that the source of these particles
with energy $\gtrsim 10^{19}\,$eV are powerful extragalactic
astrophysical objects. The candidates include active galactic nuclei
(AGN, e.g, \citealt{Biermann87, Takahara90, RB93,Berezinsky06,
  Dermer09}), gamma--ray bursts (GRBs, e.g., \citealt{Waxman95,
  Vietri95, Dermer06, Murase06}), semi--relativistic hypernova
\citep[e.g.][]{Wang07,Budnik08, Chakraborti11, Liu12a} and
extragalactic rotation-powered young pulsars
\citep{Arons03,Fang12}. 

On their way to the detector, UHECRs suffer inevitable interactions
with the cosmic microwave background (CMB) and the extragalactic
background light (EBL) that permeate extragalactic space, in
particular Bethe--Heitler pair production, photo-pion production and
photodisintegration. Photo-hadronic interactions inevitably introduce
a high energy cut--off in the UHECR spectrum beyond $\sim 5\times
10^{19}$eV\citep{Greisen66, Zatsepin66, Puget76} [GZK], due to the
rapid decrease of the attenuation length of UHECRs with increasing
energy. Above these GZK energies, the typical horizon to which sources
can be detected shrinks to values of the order of $\sim
100-200\,$Mpc. This cut--off/suppression feature has now been observed
by different experiments at a statistically significant level
\citep{Hires08,PAO08,PAO10b}, implying notably that the sources of the
highest energy cosmic rays must be nearby luminous objects
\citep[e.g.][]{Waxman95, Waxman05, Farrar09, Piran10, Taylor11a}.

The Pierre Auger Observatory (PAO), presently the largest UHECR
observatory, has reported the detection 69 events within the energy
range 55--142\,EeV between January 2004 and December 2009 \citep{PAO10a}. 
A detailed analysis has shown that the fraction of these events
correlating with nearby AGN ($<75\,$Mpc) in the V\'eron-Cetty and
V\'eron (VCV) catalog is ($38^{+7}_{-6}$)\%, above the isotropic
expectation of 21\%. Most of this excess is found around the direction
of Centaurus A (Cen A) within a surrounding 18$^{\circ}$ window, in
which 13 events in the energy range 55--84\,EeV are observed while
only 3.2 are expected \citep{PAO10a}. However, both the intergalactic
and the Galactic magnetic field deflect the trajectories of cosmic
rays, resulting in apparent correlations with objects which are not
necessarily their true birth places. Furthermore, measurements on the
maximum air shower elongations $\langle X_{\rm max}\rangle $ and their
rms ($\sigma_{X_{\rm max}}$) by the PAO suggest that the chemical
composition of UHECRs are progressively dominated by heavier nuclei at
energies above 4\,EeV \citep{PAO10c}. If the cosmic rays are indeed
intermediate--mass or heavy nuclei, the deviation of their arrival
directions due to propagation in the intervening magnetic fields must
be significant, hence the observation of anisotropies appears slightly
surprising. From a theoretical point of view, it may appear more
favorable to accelerate heavy nuclei, as their higher charge,
comparatively to protons, reduce the energetic constraints placed on
the source candidates, e.g. \citet{Lemoine09}. However, it also
requires the acceleration site to be abundant in intermediate--mass or heavy
elements, well beyond a typical galactic composition \citep{PAO11,
  Liu12b}. Finally, other experiments, such as HiRes or the Telescope
Array, find that the composition at $\gtrsim 10^{19}\,$eV remains
dominated by light nuclei \citep{Hires10, Tsunesada11}.

One way to make progress is to use the pattern of anisotropies as a
function of energy. This idea was first proposed in \citet{Lemoine09}:
if a source produces an anisotropy signal at energy $E$ with cosmic
ray nuclei of charge $Z$, it should also produce a similar anisotropy
pattern at energies $E/Z$ via the proton component that is emitted
along with the nuclei, given that the trajectory of cosmic rays within
a magnetic field is only rigidity--dependent. It is easy to show that
the low energy anisotropy should appear stronger, possibly much
stronger than the high energy anisotropy (assuming a chemical
composition similar to that inferred at the source of Galactic cosmic
rays), offering means to constrain the chemical composition of the
source. This test has been applied on the Pierre Auger Observatory
dataset and no significant anisotropy has been found at energies
$E/Z$, with $E=55\,$EeV and $Z=6,14,26$ \citep{PAO11}.

In the present work, we push further and generalize this idea by
considering the amount of secondary protons produced through
photodisintegration interactions of the primary nuclei. We
provide detailed analytical and numerical estimates of the ratio of
significance of the anisotropy at $E/Z$ vs $E$, and derive the maximal
distance to the source $D_{\rm max}$ in order to avoid the formation
of a stronger anisotropy pattern produced by the secondary protons at
energy $E/Z$. This bound does not depend on the amount of protons produced by
the source. We also discuss how the comparison of the anisotropy ratio
constrains the metal abundance in the source, independently of the
injection spectral index, and emphasize how large this metal abundance
must be, if the anisotropies persist at high energies, but not at low
energies. Finally, we briefly discuss the prospects for the detection
of anisotropies at higher energies than $E$, with next generation
experiments, based on current reports of anisotropies at $E$.

The layout of this paper is arranged as follows. In \S ~\ref{sec:aniso},
we discuss how the absence of a low--energy anisotropy signal could
constrain the source distance and its metal abundance. In \S
\ref{sec:disc}, we discuss the low energy proton fraction and the
possible anisotropy signal at higher energies, as well as their
implication for the source. We draw some conclusions in \S
\ref{sec:conc}.

\section{Anisotropies at constant rigidity}\label{sec:aniso}

\subsection{Low energy anisotropy signal}
We assume that some anisotropy is detected within a solid angle
$\Delta\Omega$ in the energy range from $E_1$ to $E_2$.  Following up
on \citet{Lemoine09}, we quantify the significance of the anisotropy
through its signal--to--noise ratio. Before doing so, we define the
injected spectrum of an element with nuclear charge number $Z$ as
\begin{equation}
q_{Z,\rm inj}=k_ZE^{-s}\ ,
\end{equation}
with $s$ the power law index and $k_Z$ the relative abundance of this
element at a given energy. Provided that the maximum energy $E_{Z,\rm
  max}$ and the minimum energy $E_{Z, \rm min}$ of the accelerated
spectrum are proportional to $Z$, i.e. scale with rigidity, the total
mass of the element of charge $Z$ and mass $A_Z$ scales as
\begin{equation}
  M_Z\propto A_Z\int q_{Z,\rm inj}dE\propto k_ZA_ZZ^{1-s}\ .
\end{equation}
Note that the above result does not depend on the magnitude of $s$,
and the missing prefactor does not depend on $Z$.  This implies in
particular that the ratio of the relative abundance of a species
at a given energy to that of hydrogen takes the form
\begin{equation}
  \frac{k_Z}{k_p}\,=\,Z^{s-1}A_Z^{-1} \frac{M_Z}{M_{\rm H}}.
\end{equation}

We then denote respectively the number of injected and propagated
primary cosmic rays with nuclear charge $Z$ in the
energy range $[E_1,E_2]$ by $N_{Z,\rm inj}(E_1;E_2)$ and $N_{Z,\rm
  prop}(E_1;E_2)$. These two quantities are related through
\begin{equation}\label{NZ1st}
  N_{Z,\rm prop}(E_1;E_2)=f_{Z,\rm surv}(E_1;E_2) N_{Z,\rm inj}(E_1;E_2),
\end{equation}
where 
\begin{equation}
  f_{Z,\rm surv}(E_1;E_2)\,\equiv\, \frac{\int_{E_1}^{E_2}
    q_{Z,\rm prop}(E)\,{\rm d}E}{\int_{E_1}^{E_2} q_{Z,\rm
      inj}(E)\,{\rm d}E}
\end{equation} is the
surviving fraction of primaries after propagation. 

As we do not know the precise composition of cosmic ray events
constituting the anisotropy signal, in a first scenario (A) we regard
the fragments with less than $Z/4$ lost nucleons as primaries ({ i.e.,} 
$q_{Z,\rm prop}=\sum\limits_{i=3/4Z}^{Z}q_{i,\rm prop}$). This
ad-hoc choice guarantees that all arriving nuclei in the energy range
$[E_1,E_2]$ which have suffered at most $Z/4$ photodisintegration
interactions retain a similar rigidity, and thus follow a similar path
in the intervening magnetic fields.  Lighter nuclei, i.e. those that
have suffered more than $Z/4$ interactions and arrive in $[E_1,E_2]$
carry higher rigidity. Depending on the intervening magnetic fields,
such cosmic rays may or may not contribute to the anisotropies, since
the magnetic fields may form a blurred image centered on the source
(with higher rigidity cosmic rays clustering closer to the source
direction), or impart a systematic shift in the arrival directions, in
which case the higher rigidity particles might lie outside
$\Delta\Omega$, e.g. \citet{WME96, KL08}.  To account for this
uncertainty, we will consider in the following an alternative scenario
(B), in which $E_2\rightarrow +\infty$ and as many photodisintegration
interactions are allowed ({ i.e.,} $q_{Z,\rm prop}=\sum\limits_{i=2}^{Z}q_{i,\rm prop}$), 
provided the nucleus arrives with energy
$>E_1$. In this scenario (B), we thus sum up over all rigidities in
excess of $E_1/Z$.

We adopt the premise that anisotropy in the arrival distribution of
UHECR nuclei has been detected at high energies between $E_{1}$ and $E_{2}$,
ie. at energies of the order of the GZK energy. Since protons with the 
same rigidity have energies between $E_1/Z$ to $E_2/Z$, one may safely 
neglect their subsequent energy losses given that their loss lengths
at these energies are of the order of $\sim1\,$Gpc, considerably larger 
than the source distance considered here ($\lesssim 100\,$Mpc). 
Photodisintegration interactions of 
nuclei with energy in the range $(A_Z/Z)[E_1,E_2]\,\approx\,[2E_1,2E_2]$ produce 
secondary protons with energy in the range $[E_1/Z,E_2/Z]$, with number
\begin{equation}\label{Np2nd}
  N_{p,\rm dis}(E_1/Z;E_2/Z)\,=\,A_Zf_{Z,\rm loss}(2E_1;2E_2)N_{Z,\rm
    inj}(2E_1;2E_2)
  \,=\,2^{1-s}A_Zf_{Z,\rm loss}(2E_1;2E_2)N_{Z,\rm inj}(E_1;E_2)\ ,
\end{equation}
where 
\begin{equation}
  f_{Z,\rm loss}(2E_1;2E_2)\,\equiv\,
  \frac{\int_{E_1/Z}^{E_2/Z}\,q_{p,\rm dis}\,{\rm d}E}
  {A_Z\int_{2E_1}^{2E_2} q_{Z,\rm inj}\,{\rm d}E}\ .
\end{equation}
In this expression, $q_{p,\rm dis}$ represents the spectrum of
secondary protons produced during propagation. 

At low energies, the primary protons also contribute to the
anisotropy, with
\begin{equation}\label{Np1st}
  N_{p,\rm prop}(E_1/Z;E_2/Z)\,\simeq\,N_{p,\rm
    inj}(E_1/Z;E_2/Z)\,=\,\frac{M_{\rm H}}{M_Z}A_ZN_{Z,\rm
    inj}(E_1;E_2)\ .
\end{equation}
Note that the last equality is of particular interest. It
shows that $N_{p,\rm prop}(E_1/Z;E_2/Z)/N_{Z,\rm inj}(E_1;E_2)=
(M_H/M_Z)A_Z$ controls the scaling of the signal-to-noise ratio 
of the low energy to high energy anisotropy signals.
This scaling factor does not depend on the injection spectrum index, but does depend
on the metal abundance at the source. It remains valid for general injection
spectra, provided this spectrum is shaped by rigidity, i.e. $q_{Z,\rm
  inj}(E) \propto \phi(E/Z)$, with $\phi$ an arbitrary function.

The noise is given by the square root of number of events expected
from the averaged all-sky spectrum of UHECRs in the same solid angle
$\Delta\Omega$. The observed spectrum of the isotropic background can be
approximately described by a broken power law beyond $\sim 10^{18}$~eV
\citep{PAO10a}, i.e.,
\begin{equation}
  \frac{{\rm d}N_{\rm iso}}{{\rm d}E}\bigg|_{\rm iso}=N_0\times \left \{ 
\begin{array}{ll}
  \left(E/E_b\right)^{-p_1} & E < E_{\rm b}, \\
  \left(E/E_b\right)^{-p_2} & E \geq E_{\rm b}
\end{array}
\right.
\end{equation}
where $p_1=2.6$, $p_2=4.3$ and $E_{\rm b}=10^{19.46}$eV. $N_0$
represents the overall amplitude, which cancels out in the following
calculation. The noise counts in the energy range $[E_1,E_2]$ then
reads
\begin{equation}
  N_{\rm iso}(E_1;E_2)\,=\,\Delta\Omega\int_{E_1}^{E_2}\frac{{\rm d}N_{\rm iso}}{{\rm d}E}\,{\rm d}E
\end{equation}
and for the low energy noise we have
\begin{equation}
  N_{\rm
    iso}(E_1/Z;E_2/Z)\,=\,\Delta\Omega\int_{E_1/Z}^{E_2/Z}\frac{{\rm
      d}N_{\rm iso}}{{\rm d}E}\,{\rm d}E=\eta
  Z^{p_1-1}N_{\rm iso}(E_1;E_2)\ ,
\end{equation} 
with $\eta\equiv
(1-p_2)(1-p_1)^{-1}\left(E_2^{1-p_1}-E_1^{1-p_1}\right)\left(E_2^{1-p_2}-E_1^{1-p_2}\right)^{-1}E_{\rm
  b}^{p_1-p_2}$. { The above equation is valid for $E_1>E_b$ and $E_2/Z<E_b$, which is the case of the "Cen A excess". If $E_2<E_b$ or $E_1/Z>E_b$,  Eq.~(11) will read $Z^{p_1-1}N_{\rm iso}(E_1;E_2)$ or $Z^{p_2-1}N_{\rm iso}(E_1;E_2)$ respectively.}

Assuming the anisotropy is mainly caused by cosmic ray nuclei with
charge $Z$, the signal-to-noise ratio in the energy range $[E_1,E_2]$
can then be expressed as
\begin{equation}\label{Sigz}
  \Sigma_Z(E_1;E_2)\,=\,\frac{N_{Z,\rm prop}(E_1;E_2)}{\sqrt{N_{\rm iso}(E_1;E_2)}}=\frac{f_{Z,\rm surv}(E_1;E_2)N_{Z,\rm inj}(E_1;E_2)}{\sqrt{N_{\rm iso}(E_1;E_2)}},
\end{equation}
while the S/N of the low energy anisotropy produced by protons with the same rigidity is
\begin{equation}\label{Sigp}
  \Sigma_p(E_1/Z;E_2/Z)\,=\,\frac{N_{p,\rm prop}(E_1/Z;E_2/Z)+N_{p,\rm
      dis}(E_1/Z;E_2/Z)}{\sqrt{N_{\rm
        iso}(E_1/Z;E_2/Z)}}\,=\,\frac{A_Z\left[M_{\rm
      H}/M_Z\,+\,2^{1-s}f_{Z,\rm loss}(2E_1;2E_2)\right]N_{Z, \rm
      inj}(E_1;E_2)}{\sqrt{\eta Z^{p_1-1}N_{\rm iso}(E_1;E_2)}}\ .
\end{equation}
Consequently, the ratio of the signal-to-noise ratios at low to high
energy reads
\begin{equation}\label{constraint2}
  \frac{\Sigma_p}{\Sigma_Z}=\frac{2M_H/M_Z\,+\,2^{2-s}f_{Z,\rm
      loss}(2E_1;2E_2)}{f_{Z,\rm surv}(E_1;E_2)\sqrt{\eta
      Z^{p_1-3}}}\ ,
\end{equation}
and if no anisotropy is recorded at low energies, one requires
$\Sigma_p/\Sigma_Z <1$. For reference, the Pierre Auger Observatory 
data indicate that, for the Cen~A excess,
$\Sigma_p/\Sigma_Z\,\lesssim\,(2.,1.8,0.8)$ at the 95\% c.l., for
$Z=6,14,26$ corresponding to carbon, silicon and iron \cite{PAO11}. The
exact number depends on the statistics (which of course have increased
since this analysis was carried out), and on the elements adopted in the 
analysis. In the following, we use the constraint $\Sigma_p/\Sigma_Z <1$
to impose a limit on the maximum source distance.

In terms of the (inverse) metal abundance, this constraint can be
rewritten
\begin{equation}\label{final_LW}
  \frac{M_{\rm H}}{M_Z}<\frac{1}{2}\left[\sqrt{\eta Z^{p_1-3}}f_{Z,\rm
      surv}(E_1;E_2)-2^{2-s}f_{Z,\rm loss}(2E_1;2E_2)\right]\ .
\end{equation}
Again, if secondary protons are ignored, meaning $f_{Z,\rm
  loss}(2E_1;2E_2)\rightarrow 0$ in the above, then the non-detection
of anisotropy at low energies imposes a lower limit of  $M_Z/M_{\rm H}$
which does not depend on the spectral index.  Note that this statement is not in contradiction with the statement in the \citet{PAO11}, that the limit on the quantity $f_p/f_Z$ used in that paper depends on the spectral index. This is due to the fact that $f_p/f_Z$, which is defined at a given energy, is not the "proton to heavy fraction in the source" or the "relative proton abundance", as it is misleadingly referred to in the Auger paper (in the notation of that paper, the relative proton abundance is $k_p/k_Z$, which is equivalent to our $M_p/M_Z$).

The minimum required metallicity of the element responsible for the
observed anisotropy thus depends on the value of $f_{Z,\rm surv}$ and
$f_{Z,\rm loss}$, which are directly determined by the source
distance. A larger source distance will result in a smaller $f_{Z,\rm
  surv}$ and a larger $f_{Z,\rm loss}$ as more nuclei are
photodisintegrated. There exists therefore a critical distance, beyond
which the abundance of hydrogen in the source relative to metals
becomes negative. This happens when $f_{Z,\rm
  loss}(2E_1;2E_2)/f_{Z,\rm surv.}(E_1;E_2)\geq 2^{s-2}\sqrt{\eta
  Z^{p_1-3}}$, meaning that even if the source injects no primary
protons, secondary protons produced during propagation cause a
stronger anisotropy at low energies. Therefore, the critical distance,
which we denote by $D_{\rm max}$ hereafter, is the upper limit of the
distance of the source responsible for the anisotropy signal in
$[E_1,E_2]$. The value of $D_{\rm max}$ is also related to the primary 
cosmic ray species adopted and the injection spectrum used. Once these
parameters are given, we can uniquely determine $D_{\rm max}$ by finding the distance
for which $\Sigma_p(E_1/Z;E_2/Z)/\Sigma_Z(E_1;E_2)=1$ using the method
outlined above.

\subsection{Photodisintegration of nuclei}
Ultra-high energy cosmic rays interact with the CMB and EBL photons
while they propagate through extragalactic space. For nuclei, energy
losses due to the photodisintegration process and the Bethe--Heitler
process (pair production) by CMB photons are comparable around
55\,EeV. As energy increases, the photodisintegration process plays a
more and more dominant role in the energy loss process. Photodisintegration 
does not change the Lorentz factor of the cosmic ray nucleus, but does lead 
to the nucleus losing one or several nucleons as well as
$\alpha$ particles through the giant dipole resonance (GDR) or
quasi--deuteron (QD) process. These secondary nuclei can be further
disintegrated to protons. On average, the mass number of a nucleus
evolves as \cite{Stecker69}
\begin{equation}
  -\frac{{\rm d}A}{{\rm d}x}\,=\,\frac{1}{2\gamma_A^2}\sum_i \Delta A_i
  \int_{\epsilon_{{\rm th},i}}^{\infty}{\rm d}\epsilon\, \sigma_{{\rm dis},i}^A
  (\epsilon)\epsilon\,\int_{\epsilon/2\gamma_A}^{2\gamma_A\epsilon}{\rm
    d}\epsilon_{\gamma}\,
  \frac{n_{\gamma}(\epsilon_{\gamma},z)}{\epsilon_{\gamma}^2}
\end{equation}
with $\sigma_{{\rm dis}, i}^A$ the cross-section for
photodisintegration through the $i$th channel (e.g., single--nucleon
emission, deuterium emission, $\alpha$ particle emission and so on),
and $\epsilon_{{\rm th},i}$  the threshold energy of the $i$th
channel, which is $\sim 10-20\,$MeV for all species of nuclei for the
GDR process and $\sim 30\,$MeV for the QD process. $\Delta A_i$ is the
number of nucleon lost through the $i$th channel (e.g., $\Delta
A=1,2,4$ for single nucleon emission, deuteron and $\alpha$ particle
emission and so on). $\epsilon_\gamma$ and
$n_{\gamma}(\epsilon_{\gamma},z)$ are respectively the photon energy
and the number density of the target photon field in the lab frame at
redshift $z$ while $\epsilon$ is the photon energy in the rest frame
of the nucleus. The physics of UHE nuclei transport through the
radiation backgrounds has been discussed by a number of authors,
e.g. \citet{Puget76, Bertone02, Khan05, Hooper08, Aloisio12}. In this
work, we will adopt the tabulated cross-section data generated by the
code TALYS and implement them into the Monte--Carlo framework along
with other energy loss processes, as described in \cite{Hooper07}, to
obtain the propagated spectra.

\begin{figure}[ht]
  \plotone{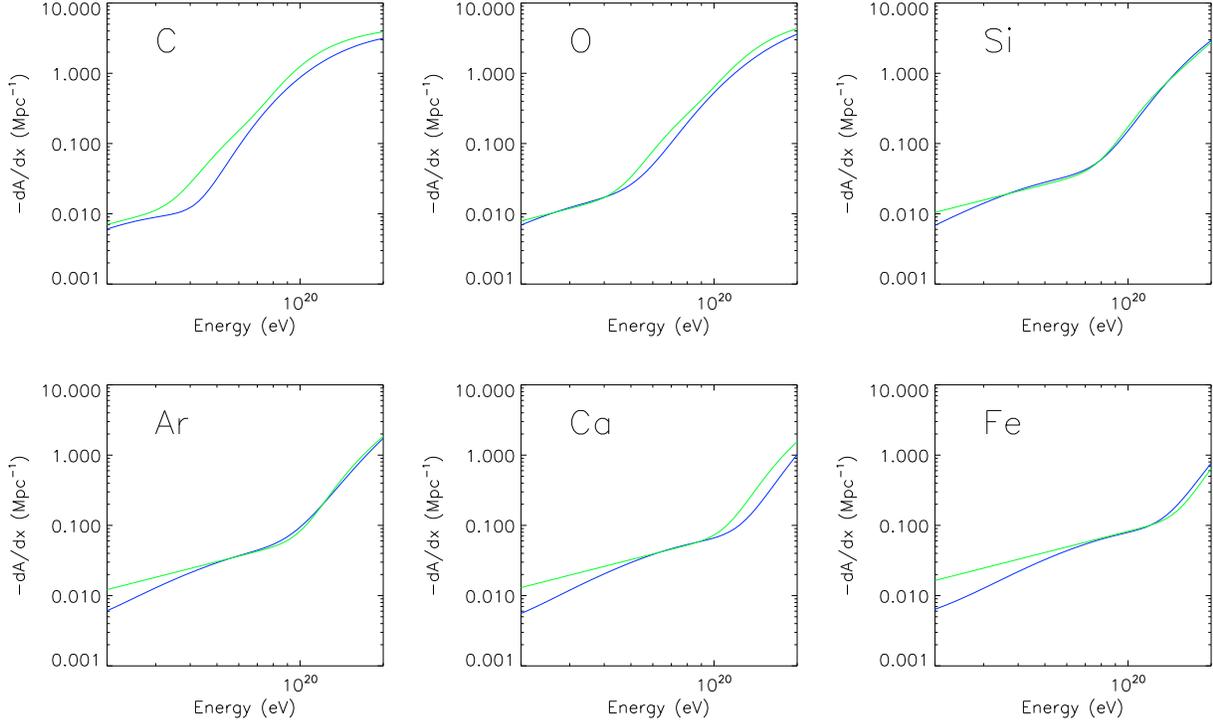}\caption{The phenomenological fit of
    nucleon loss rate for some species of cosmic rays. The blue lines
    are numerically calculated results while the green ones are
    plotted based on the analytical expressions
    (Eq.~\ref{dA2dx}).\label{LossRateFit}}
\end{figure}

As Monte Carlo simulations of nuclei propagation remain somewhat
costly in computing, it is useful to have a simple analytical
estimate of the photodisintegration process. Detailed treatments are
discussed in \citet{Hooper08,Aloisio13a,Aloisio13b}. Here, we adopt an
even simpler approximation, which provides a sufficiently reliable
approximation for the integrations that follow. In this analytical
treatment, only the photodisintegration process is taken into account,
with all other energy loss processes being neglected. A phenomenological 
fit of the nucleon loss rate ${\rm d}A/{\rm d}x$ for a nucleus with initial 
mass number $A_0$ and Lorentz factor $\gamma$:
\begin{equation}\label{dA2dx}
  -\frac{{\rm d}A}{{\rm d}x}=c_1(\gamma_{10})A^2+c_2(\gamma_{10})A~~\rm Mpc^{-1}
\end{equation}
where $c_1(\gamma_{10})$ and $c_2(\gamma_{10})$ are functions of the
Lorentz factor of cosmic ray nuclei in unit of $10^{10}$, which can be
written in the form of $a_1\gamma_{10}+a_2{\rm
  exp}(-a_3/\gamma_{10}^{a_4})$.

\begin{table}
\begin{center}
\caption{The best-fit parameters for nucleon loss rate through a MCMC method. \label{tbl-1}}
\begin{tabular}{ccc}
\tableline\tableline
~ & $c_1$ & $c_2$\\
\tableline
$a_1$ & $9.99\times 10^{-5}$ & $2.11\times 10^{-3}$ \\
$a_2$ & $7.43\times 10^{-3}$ & $0.31$ \\
$a_3$ & $0.69$ & 1.15 \\
$a_4$ & $1.79$ & 2.80 \\
\tableline
\end{tabular}
\end{center}
\end{table}

Table~1 presents the results from a Markov Chain Monte Carlo
exploration of the parameter space and Fig.~\ref{LossRateFit} shows
the phenomenologically fit and numerical nucleon loss rate
respectively for several cosmic ray species. From Eq.~(\ref{dA2dx}) we
can derive the average mass number of a nucleus of initial mass number
$A_0$ and Lorentz factor $\gamma_{10}$ after propagation over a distance
$x$:
\begin{equation}\label{Ax}
A(x,\gamma_{10})=\frac{A_0c_2e^{-c_2x}}{A_0c_1(1-e^{-c_2x})+c_2}\ .
\end{equation}
For Poisson statistics with mean rate ${\rm d}A/{\rm d}x$, the
probability that a nucleus undergoes at most $N$ interactions reads
\begin{equation}
P_N(A_Z,x,E)\,\equiv\,\frac{\Gamma\left[N+1,x\,|{\rm d}A/{\rm
      d}x|\right]}{\Gamma\left[N+1\right]}.
\end{equation}
Strictly speaking, Eq.~(\ref{dA2dx}) leads to modified Poisson
statistics, because the rate ${\rm d}A/{\rm d}x$ depends on $A$, which
evolves as photodisintegration interactions occur. It is possible to
derive the generalized probability law for $P_N$, at the expense of
tedious calculation; however, as we demonstrate in the 
following, the above form for $P_N$ provides a sufficient approximation 
for our case of interest.

One may then derive the propagated spectra, surviving fractions and
secondary proton spectra used in the previous subsection, as
follows. Consider first the simpler scenario (B) in which one sums up
over all fragments with rigidities in excess of $E_1/Z$. Writing
$Q_{Z,\rm prop}(E)=E\,q_{Z,\rm prop}(E)$ the number of particles per
log interval, and neglecting losses other than photodisintegration,
one finds
\begin{equation}
  Q_{Z,\rm prop}(E)\,=\, \sum_{i=0}^{i=A_Z-1}\,p_i\left(A_Z,x,\frac{E}{1-i/A_Z}\right) Q_{Z,\rm
    inj}\left(\frac{E}{1-i/A_Z}\right)\ ,\label{eq:eqz}
\end{equation}
with $p_i(A_Z,x,E)\,\equiv\,\left(x\,|{\rm d}A/{\rm d}x|\right)^i
\exp\left(-x\,|{\rm d}A/{\rm d}x|\right)/i!$ the probability to undergo
$i$ photodisintegration interactions over a distance $x$, thereby
decreasing the injection energy from $E/(1-i/A_Z)$ down to $E$.

Then, the fraction of surviving fragments with rigidity $>E_1/Z$ can
be obtained as
\begin{equation}
  f_{Z,\rm surv}(>E_1)\,=\,\frac{1}{\int_{E_1}^{+\infty}\,q_{Z,\rm
      inj}(E)dE}\,\int_{\ln E_1}^{+\infty}{\rm d}\ln E\,Q_{Z,\rm
    prop}(E)\,=\,
  \frac{1}{\int_{E_1}^{+\infty}\,q_{Z,\rm
      inj}(E)dE}\,\int_{E_1}^{+\infty} {\rm d}E\,q_{Z, \rm
    inj}(E)\,P_j(A_Z,x,E)\ ,
\end{equation}
with $j\,\equiv\, {\rm min}\left\{A_Z-1,{\rm Int}[A_Z\left(1-E_1/E\right)]\right\}$. The latter equality is
obtained by inverting the summation interval between the integral and
the discrete sum, { changing variables in the integration from $E \rightarrow E/(1-i/A)$, then permuting the order of integration. Note that $P_j(A_Z,x,E)=\sum_{i=0}^{j}p_i(A_Z,x,E)$}.

The fraction of photodisintegrated nuclei with energy more than $2E$ is given by $f_{Z,\rm
  loss}(>2E)=1-f_{Z,\rm surv}(>2E)$, and the number of secondary
protons is easily evaluated using Eq.~(\ref{Np2nd}).

In Scenario (A), one considers only the fragments with energy in
the range $[E_1,E_2]$, which have suffered at most $Z/4$ photodisintegration
interactions, so as to study a group of nuclei with similar
rigidities. Eq.~(\ref{eq:eqz}) remains valid, if the sum over $i$ runs
from $i=0$ to $i=Z/4$, therefore one finds 
\begin{equation}
f_{Z,\rm surv}(E_1;E_2)\,=\,\frac{1}{\int_{E_1}^{E_2}\,q_{Z,\rm
      inj}(E)\,{\rm d}E}\left\{\int_{E_1}^{+\infty} dE q_{Z, \rm
      inj}(E)\,P_{j_1}(A_Z,x,E)- \int_{E_2}^{+\infty} dE q_{Z, \rm
      inj}(E)\,P_{j_2}(A_Z,x,E)\right\}\ ,
\end{equation}
with $j_1={\rm min}\left\{Z/4,{\rm int}[A\left(1-E_1/E\right)]\right\}$, $j_2={\rm
  min}\left\{Z/4,{\rm int}[A\left(1-E_2/E\right)]\right\}$. Here as well, one
defines $f_{Z,\rm loss}(>2E)=1-f_{Z,\rm surv}(>2E)$, and the number of
secondary protons is easily evaluated using Eq.~(\ref{Np2nd}).

\subsection{Results}

\begin{figure}[ht]
  \plotone{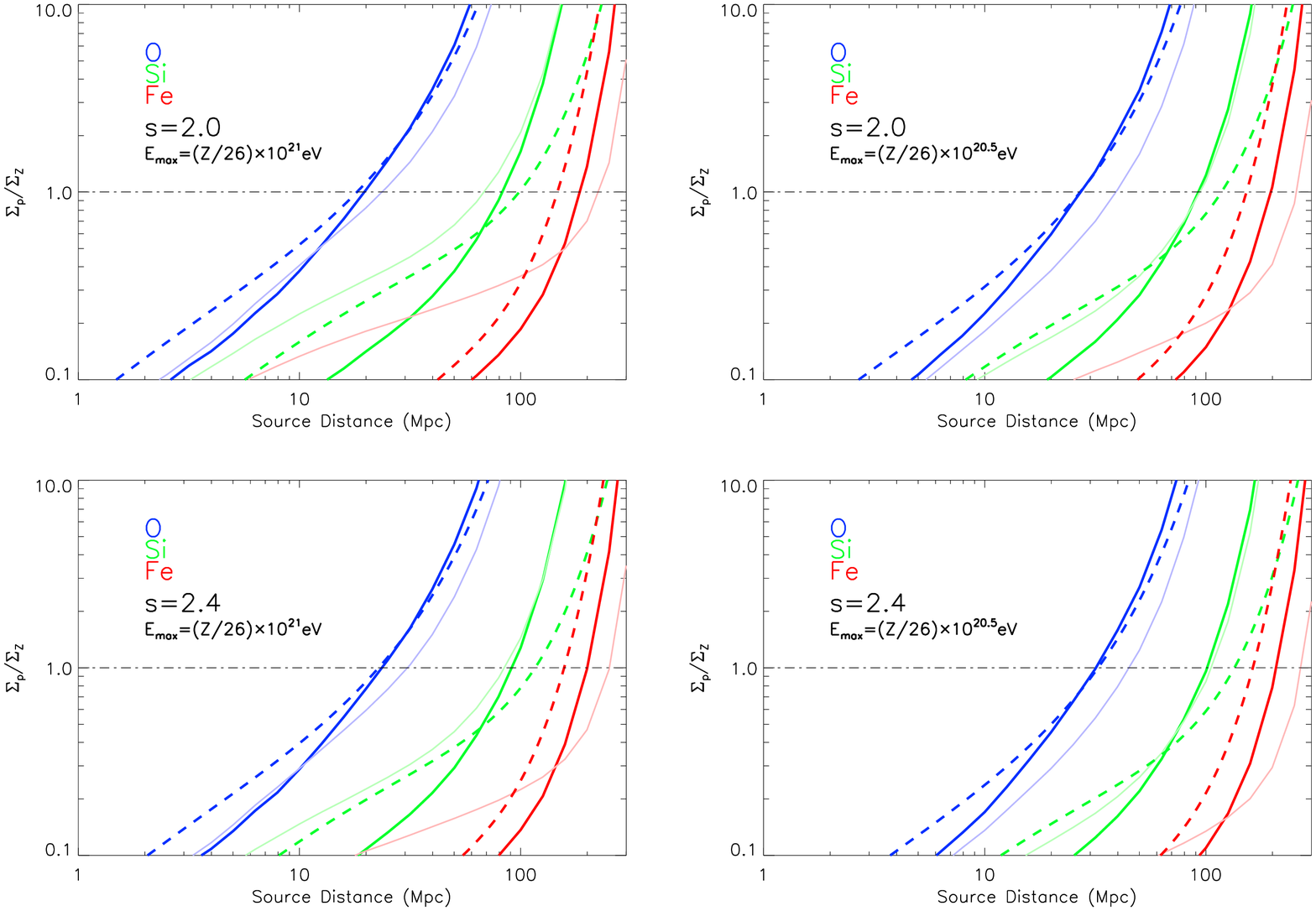}\caption{Ratio of anisotropy significance at
    low to high energies as a function of the distance to the sources
    responsible for the anisotropy. Solid lines represent the
    numerical results while dashed lines represent the analytical
    results; thick solid line: scenario (A), in which one sums up over
    fragments of similar rigidity, in interval $[E_1,E_2]$, with at
    most $Z/4$ photodisintegration interactions; thin solid lines:
    scenario (B), in which one sums up over all fragments with
    rigidities in excess of $E_1/Z$. The source is assumed to inject
    pure O, Si or Fe composition as indicated. \label{snratio}}
\end{figure}

So far, our treatment has remained quite general. Here, we apply it to
the specific case of the "Cen A excess" reported by the Pierre Auger
collaboration. We thus use $E_1=55\,$EeV, $E_2=84\,$EeV and assume for
simplicity that the source injects a pure oxygen, silicon or iron
composition.  In Fig.~\ref{snratio} we show both the analytical and
the numerical results of the ratio of anisotropy significance at low
to high energies as a function of the distance to the source that is
responsible for the anisotropy. In this figure, we do not assume any
proton component in the source composition, so that $N_{p,\rm prop}
\rightarrow 0$, $M_{\rm H}/M_Z\rightarrow 0$ in Eqs.~(\ref{Sigp}) and
(\ref{constraint2}).  We adopt an exponential cut--off power law
spectrum, as generally expected, with a cut--off energy $E_{\rm max} \propto
Z$. The four panels correspond to different injection spectral indices
and maximal energies. As can be seen, these results share the
following common features.

At small source distances, the anisotropy signal produced by secondary
protons is less prominent than the high energy one, because only a few
primary nuclei photodisintegrate on this short path length. As more 
and more secondary protons are produced with increasing source distance,
this ratio grows and eventually exceeds unity. In the case $s=2,
E_{\rm max}=(Z/26)\times 10^{21}\,$eV, both the analytical treatment
and the numerical treatment result in a maximum source distance of
$\sim 15$\,Mpc, $\sim 60$\,Mpc and $\sim 180\,$Mpc for oxygen nuclei,
silicon nuclei and iron nuclei respectively. Note that for $E_{\rm
  max}=(Z/26)\times 10^{21}\,$eV, the source produces protons of
energy $\gtrsim 40\,$EeV: such protons would presumably produce a
strong anisotropy, though its magnitude would depend strongly on the
distribution and characteristics of intervening magnetic fields. We
show therefore the case with a high $E_{\rm max}$ for the sake of
generality in order to illustrate the dependence of the results on the
maximal energy.

From lighter to heavier nuclei, the constraint on the source
distance becomes weaker, since at a given energy lighter nuclei carry
a comparatively larger Lorentz factor, and as a consequence, their
energy lies further beyond the photodisintegration interaction threshold 
(see also Fig.~\ref{LossRateFit}). The small differences
of $D_{\rm max}$ for the same species among the four panels can be
interpreted as follows: protons at $E/Z$ all come from primary nuclei
at $2E$, so a smaller cut--off energy or a steeper power-law slope
will decrease the amount of primary nuclei at $2E$, leading to less
secondary protons produced at $E/Z$, so that the values of maximum
source distances in these cases are larger.

\begin{figure}[ht]
  \plotone{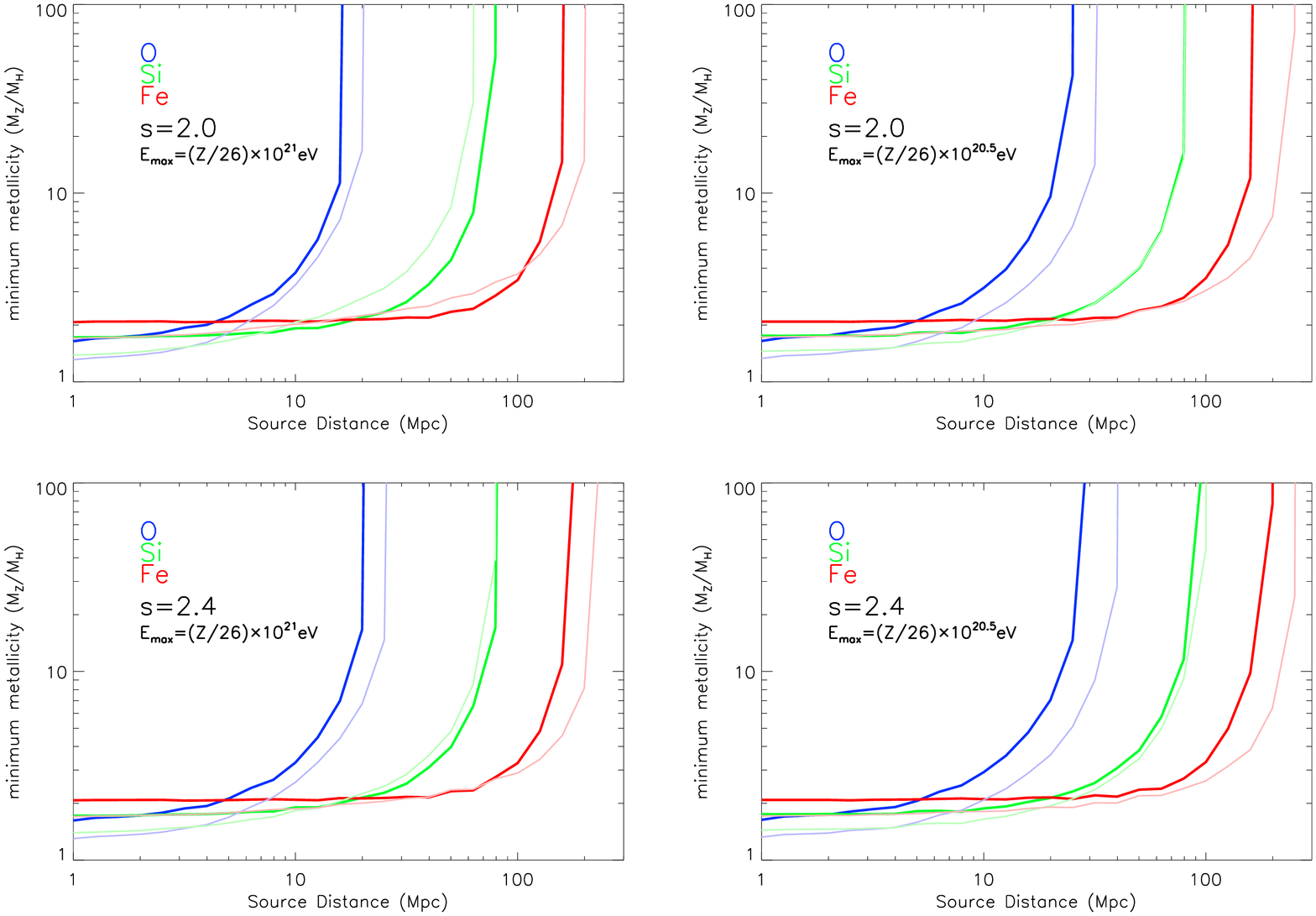}\caption{The minimum metal mass relative to
    hydrogen in the source, assuming pure O, Si or Fe compositions
    are injected. Thick solid lines and thin solid lines respectively
    represent results in scenario (A) and (B), which are the same as
    in Fig.~\ref{snratio}. \label{metallicity}}
\end{figure}

Another way to plot these results is to consider the minimum metal
abundance $M_Z/M_{\rm H}$ that is required at the source in order to
satisfy the bound $\Sigma_p/\Sigma_Z<1$. The results are shown in
Fig.~\ref{metallicity} as a function of the distance to the
source. All four panels indicate that the mass ratio of nuclei to
proton $\gtrsim 1:1$ is needed. Of course, as the distance increases,
so does the minimum $M_Z/M_{\rm H}$, in order to compensate for the
greater number of secondary protons produced during propagation.  The
distance where $M_Z/M_H\rightarrow +\infty$ corresponds to $D_{\rm
  max}$. Conversely, the asymptote as $D\rightarrow 0$ indicate the
minimum $M_Z/M_{\rm H}$ amount when secondary protons can be safely
neglected.

\section{Discussion}\label{sec:disc}
We emphasize the method that we have presented remains quite general
and could be applied to datasets of next generation
experiments. Nevertheless, the results obtained in
Figs.~\ref{snratio},\ref{metallicity} assume tacitly that the heavy
chemical composition and the anisotropy signal reported by the PAO are
not artifacts.  It is fair to say that these two results remain
disputed. The significance level of the anisotropy, for instance, is
not comfortably high, and deserves to be improved with extended
datasets. The measurements of the chemical composition by the High
Resolution Fly's Eye Experiment (HiRes) and Telescope Array (TA)
differ appreciably from that of the PAO. In particular, their data of
$\langle X_{\rm max}\rangle $ and rms $\sigma_{X_{\rm max}}$ show a
proton dominated spectrum at all energies
$>10^{18}\,$eV\citep{Hires10, Tsunesada11}. One should not expect to
detect anisotropies at low energies if the composition were pure
proton, as the low energy protons have a much smaller rigidity than
the high energy ones. On the other hand, the analysis of the chemical
composition depends on the details of the hadronic interaction model,
such as the cross sections, multiplicities and so on. The fact that
these parameters are poorly constrained at present prevents one from
drawing firm conclusions. As for the apparent anisotropy,
\citet{Clay10} has shown that there is no significant difference
between the energy distribution of the events inside and outside the
$25^{\circ}$ window of Cen A using a K-S test, implying events around
Cen A do not have any special origin; such an analysis cannot provide
however a conclusive answer, given the limited event statistics
presently available. Additionally, two recent papers suggest that at
most 5--6 of events around Cen A can originate from it by backtracing
the events' trajectories in the intervening magnetic field
\citep{Farrar12, Sushchov12}. We should, however, be cautious with
such strong conclusions given that they depend on the magnetic field
model adopted, which still carries a large degree of uncertainty.

\subsection{Source metallicity}
With the above caveats in mind, it is interesting to discuss where the
previous results lead us. The constraints derived from
Fig.~\ref{metallicity} are indeed quite strong. For reference, the
solar composition \citep{Lodders09} corresponds to $M_{\rm H}/M_{\rm CNO}\,\sim\, 70$,
$M_{\rm H}/M_{\rm Si}\,\sim \, 900$ and $M_{\rm H}/M_{\rm Fe
}\,\sim\,550$. Consequently, the minimum metallicities required to
match $\Sigma_p/\Sigma_Z<1$, notwithstanding the secondary protons,
are $\sim 120 Z_\odot$ for CNO, $\sim 1600Z_\odot$ for Si and $\sim
1100Z_\odot$ for iron like nuclei. The comparison to $Z_\odot$ is less
severe for oxygen, but this nucleus is also more fragile and the
minimum metallicity diverges rapidly beyond some
$20-30\,$Mpc. Conversely, the production of secondary protons is less
severe for iron nuclei, but for such nuclei, the minimum requirements
on the source metallicity are already quite extraordinary.

The observables $\langle X_{\rm max}\rangle$ and $\sigma_{X_{\rm
    max}}$, as reported by the Pierre Auger Observatory, suggest that
the all-sky-averaged composition of arriving UHECRs may be oxygen--like
\citep{Hooper10}. If the anisotropy signal observed by the PAO mainly
consists of oxygen--like nuclei, our calculations indicate that the
source responsible for the anisotropy should lie within
$20-30\,$Mpc. There are only a limited number of known powerful
radio-galaxies within this distance, such as Cen A, M87. Such
radio-galaxies are relatively weak, in terms of jet power and magnetic
luminosity, which implies that they cannot accelerate particles beyond
$E_{\rm max}\,\sim Z\times 10^{18}-10^{19}\,$eV, see the discussion in
\citet{Lemoine09}. Even assuming that these sources accelerate oxygen
nuclei to the highest energies, the minimum metallicity required by
the above arguments lies well above what is measured in the central
parts of such radio-galaxies \citep{Hamann99}. The situation becomes even
worse if one considers silicon or heavier nuclei. Consequently, and as
already emphasized in \citet{Lemoine09}, the current dataset of the
Pierre Auger Observatory, in particular the clustering towards Cen~A,
does not provide support for acceleration of UHECRs in this object.

If future datasets confirm the existence of anisotropies at high
energies, and the absence of anisotropies at low energies, then the
present work provides strong constraints on the nature and the source
of ultra-high energy cosmic rays: either protons exist at ultra-high
energies, and some of them are responsible for the observed
anisotropies (in which case no anisotropy is indeed expected at lower
energies); or, a close-by source with rather extraordinarily high
metallicity produces these anisotropies. The only physically motivated
scenario for such a source so far is acceleration at the external
shock of a semi-relativistic hypernovae inside the wind of the
progenitor \citep{Wang07,Budnik08, Chakraborti11,Liu12a}.

\subsection{Composition close to the ankle}
Provided the same source population produces both UHECRs with energy
$>E_1$ and $>E_1/Z$, the proton fraction at $E_1/Z$ becomes an
interesting aspect of the problem. The key point indeed is that if
$M_Z\,\gtrsim\,M_{\rm H}$ inside the sources, as suggested by the
above discussion, and all sources are alike, then the chemical
composition at $E_1/Z$ must contain a significant heavy
component\footnote{We thank S. Nagataki for suggesting this to us}.
More specifically, the fraction of protons at low energies is given by
\begin{equation}\label{xp}
  x_p(E_1/Z;E_2/Z)=\frac{\tilde N_p(E_1/Z;E_2/Z)}{\tilde N_{Z,\rm
      prop}(E_1/Z;E_2/Z)+\tilde N_p(E_1/Z;E_2/Z)}
\end{equation}
where $\tilde N_p=\tilde N_{p, \rm prop}+ \tilde N_{p, \rm dis}$ is the
total proton number, including the contribution from secondary protons
and primary protons, as integrated over all sources, and similarly for
$\tilde N_{Z,\rm prop}$.  Here we neglect the partially disintegrated
fragments.  Assuming every source has equal emissivity and the same
injection spectrum, we have
\begin{equation}\label{Nztot}
  \tilde{N}_{Z,\rm prop}(E_1;E_2)\,\simeq\,
  \int_{(1+z)E_1/Z}^{(1+z)E_2/Z}q_{Z,\rm inj}(E)\,{\rm
    d}E\,\int_0^{l_{Z,\rm loss}(E)} n(z)f_{Z,\rm surv}(E)\,{\rm
    d}D_{\rm c}(z)
\end{equation}
and 
\begin{equation}\label{Nptot}
  \tilde{N}_{p,\rm dis}(E_1;E_2)\,\simeq\,
  A_Z\int_{2(1+z)E_1}^{2(1+z)E_2}q_{Z,\rm inj}(E)\,{\rm d}E\int_0^{l_{p,\rm
      loss}(E/A_Z)} n(z)f_{Z,\rm loss}(E)\,{\rm d}D_{\rm c}(z).
\end{equation}
Here $n(z)$ is the source density as a function of redshift $z$ and
$D_{\rm c}(z)$ is the comoving distance to the light cone at redshift
$z$; $l_{Z,\rm loss}$ and $l_{p,\rm loss}$ represent the energy loss
lengths $\left\vert E/(dE/dx)\right\vert$ for nuclei and protons
respectively. The second equation assumes that photodisintegration
takes place on short distance scales compared to $l_{p,\rm
  loss}(E/A_Z)$, which is a very good approximation. The energy losses
to be considered here includes all processes besides
photodisintegration, such as pair production, adiabatic cooling
etc. Since the energy loss distance of protons with energy $(1+z)E/Z$
is much larger than the energy loss distance of nuclei at energy
$2(1+z)E$, $f_{Z,\rm loss}\left[2(1+z)E\right]\rightarrow 1$ for most
sources. On the other hand, $f_{Z,\rm
  surv.}\left[(1+z)E/Z\right]\,\approx\,\exp\left\{-D_{\rm c}/l_{Z,\rm
    loss}\left[(1+z)E/Z)\right]\right\}$ and given that $l_{Z,\rm
  loss}\left[(1+z)E/Z\right]$ is the upper limit of integration,
we have\footnote{If we
  also consider the slightly disintegrated fragments as surviving
  primaries, $f_{Z,\rm surv.}$ will be closer to unity} 
$e^{-1}\lesssim f_{Z,\rm surv.}\lesssim 1$. As an
estimation here we take $f_{Z,\rm surv.}=1$. Then, the above two
equations can be written as
\begin{equation}
  \tilde{N}_{Z,\rm prop}(E_1;E_2)\,\approx\, k_Z Z^{s-1}\bar{E}^{1-s}
  \int_0^{l_{Z,\rm loss}(\bar{E}/Z)} (1+z)^{1-s}n(z)\,{\rm d}D_{\rm c}
\end{equation}
\begin{equation}
  \tilde{N}_{p,\rm dis}(E_1;E_2)\,\approx\, k_Z
  2^{1-s}\bar{E}^{1-s}A_Z\int_0^{l_{p,\rm
      loss}(\bar{E}/Z)}(1+z)^{1-s}n(z)\,{\rm d}D_{\rm c}.
\end{equation}
Since $[E_1,E_2]$ is a narrow energy range, we denote the average
energy in this range by $\bar{E}$. Considering that $n(z)$ usually
evolves with redshift $z$, we make here a further approximation that
the term $(1+z)^{1-s}$ cancels the evolution in $n(z)$ to some extent
and the integrand is reduced to a constant. Then one can find that
\begin{equation}
  x_p(E_1;E_2)\,\approx\,
  \frac{1+2^{s-1}M_H/M_Z}{1+2^{s-1}M_H/M_Z\,+\,A_Z^{s-2}l_{Z,\rm
      loss}(\bar{E}/Z)/
    l_{p,\rm loss}(\bar{E}_{\rm thr}/Z)}\ .
\end{equation}
In the local Universe, for oxygen nuclei, $[E_1/Z,E_2/Z]\,\sim\,
[7,10]\,$EeV, hence the energy loss in this energy range is comparably
caused by photodisintegration on EBL photons and pair production on
CMB photons, leading to an energy loss length of $\sim 2-3\,$Gpc. For
silicon and iron nuclei, $[E_1/Z,E_2/Z]\, \sim [4,6]\,$EeV and
$[2,3]\,$EeV respectively, in which energy range the dominant cooling
process is adiabatic cooling with an energy loss length $\sim
4\,$Gpc. For protons, however, the dominant energy loss process in the
corresponding energy range is caused by pair production on CMB photons
with an energy loss length $\sim 1-2\,$Gpc. Therefore typically, the
energy loss length for nuclei is larger than that for protons by a
factor of 2-3. If $2^{s-1}\,M_{\rm H}/M_Z\,\lesssim\,1$, as suggested
by the previous discussion, this implies in turn that the composition
in $[E_1/Z,E_2/Z]$ should comprise less than $\sim 50\,$\% protons, in
potential conflict with the claims of a light composition close to the
ankle of the cosmic ray spectrum.

Looking at this argument the other way round, the data from the Pierre
Auger Observatory shows evidence for the UHECR mass composition
becoming progressively heavier at energies $\gtrsim\,$ 4\,EeV, while
below this energy the same data suggests a proton-like composition.
Thus, we should expect $M_{\rm H}/M_Z \gtrsim 1$ if the heavy elements
at $E_1$ are mostly silicon or iron nuclei. In this case,
Fig.~\ref{metallicity} indicates that one should have detected a
secondary anisotropy at the ankle.  For oxygen, however, $E_1/Z$
already lies in an energy range where the composition apparently
departs from proton-like, and the above argument is severely weakened.

\subsection{Trans-GZK anisotropies}
Another interesting aspect is the possible anisotropy signal that one
may expect at higher energies, given the reported anisotropies at
$>55\,$EeV. The detection of such anisotropies provides a strong
motivation for next generation experiments such as JEM-EUSO \citep{Casolino11}, 
which will provide a substantially larger amount of statistics.

Here we start by assuming that the currently observed anisotropy
mainly consists of nuclei with charge number $Z$ and that their
source also accelerates heavier nuclei with nuclear charge number
$Z^{'}$. These heavier nuclei will produce a similar anisotropy
pattern at higher energies $>Z'E_1/Z$; we define $E_1'=Z'E_1/Z$ for
clarity. The ratio of significance between these two anisotropy
signals then reads
\begin{equation}
  \frac{\Sigma_{Z'}(>E_1')}{\Sigma_Z(>E_1)}=\frac{M_{Z'}}{M_Z}\frac{f_{Z',\rm
      surv.}(>E_1')}{f_{Z,\rm
      surv.}(>E_1)}\left(\frac{Z^{'}}{Z}\right)^{(p_2-3)/2}\ ,\label{eq:she}
\end{equation} 
where $f_{Z,\rm surv.}\approx \exp\left(-x/l_{Z,\rm
    loss}\right)$. With the approximation $A_Z\simeq 2Z$, the two
species of nuclei with the same rigidity share approximately the same Lorentz
factor. At the same Lorentz factor, heavier nuclei lose energy faster
than lighter nuclei, but the differences between the energy loss
lengths of different species such as O, Si, Fe with rigidity $E/Z$ are
at most a factor of a few; furthermore, the energy loss lengths are
larger than the maximum source distance $D_{\rm max}$ that we have
obtained in Section 3. So we expect $0.1\,\lesssim\, f_{Z',\rm
  surv}(>E_1')/f_{Z,\rm surv}(>E_1)\,\lesssim\, 1$. Also, $p_2=4.3$,
hence $\left(Z'/Z\right)^{(p_2-3)/2}\,\gtrsim\, 1$. Therefore, a
stronger anisotropy signal is expected at higher energies if the
source is more abundant in nuclei $Z'$ than nuclei $Z$. In this case, 
however, some accompanying effects will occur and one should also check 
whether these effects already cause violations against current measurements
or lead to self-contradiction. We consider here the following three aspects.

\begin{itemize}
\item secondary protons produced by nuclei $Z'$ above energy
  $E_1/Z$. Since nuclei $Z'$ at $E_1'$ have the same rigidity as the
  nuclei $Z$ at $E_1$, the secondary protons emitted by nuclei $Z'$
  will fall well within the energy range of interest. According to
  Eq.~(\ref{Np2nd}), we can write the secondary protons above $E_1/Z$ as
  $N_{p,\rm dis}(>E_1/Z)=A_Zf_{Z,\rm loss}(>2E_1)N_{Z,\rm inj}(>2E_1)$. As
  such, we obtain the ratio between the number of secondary protons
  produced by nuclei $Z^{'}$ and $Z$ above $E_1/Z$
\begin{equation}
  \frac{N_{p',\rm dis}}{N_{p,\rm dis}}=\frac{M_{Z'}}{M_Z}\frac{f_{Z',\rm loss}(>2E_1')}{f_{Z,\rm loss}(>2E_1)}
\end{equation} 
As was discussed above, the energy loss length of nuclei $Z'$ is a bit
larger than that of nuclei $Z$ with the same Lorentz factor, so
$f_{Z',\rm loss}(>2E_1')\, \gtrsim\, f_{Z,\rm loss}(>2E_1)$. If the
source is more abundant in nuclei $Z'$ than nuclei $Z$, i.e,
$M_{Z'}>M_Z$, nuclei $Z'$ will actually produce more secondary protons
than nuclei $Z$, above energy $E_1/Z$.  As such, when we calculate the
low energy proton anisotropy significance, we should also consider the
contribution from nuclei $Z'$ and add a non--trivial term to the
numerator of Eq.~(\ref{Sigp}). Consequently, the maximum source
distance derived previously would be further reduced.

\item the chemical composition of UHECRs at energy $E_1'$: as the
  UHECR background decreases rapidly with increasing energy, the
  composition of cosmic rays emitted by the source can strongly
  influence the composition measurement at higher energies, provided
  the source accelerates a larger fraction of nuclei $Z'$ than nuclei
  $Z$. Although it is difficult to find a quantitative relation
  between the composition of the source and that of the all--sky
  averaged composition, one might naively expect that the all--sky
  averaged composition above a given energy $E$ (denoted as $\xi_Z(>E)$) to be positively
  related to $A_{Z}\frac{N_{Z,\rm prop}(>E)}{N_{\rm
      iso}(>E)}$, and we find
\begin{equation}\label{comp}
  \frac{\xi_{Z'}(>E_1')}{\xi_Z(>E_1)}=\frac{M_{Z'}}{M_Z}\frac{f_{Z',\rm surv.}(>E_1')}{f_{Z,\rm surv.}(>E_1)}\left(\frac{Z'}{Z}\right)^{p_2-1}=\left(\frac{Z'}{Z}\right)^{\frac{p_2+1}{2}} \frac{\Sigma_{Z'}}{\Sigma_Z}
\end{equation}
As one can see, since $Z'>Z$, if stronger anisotropy signal is detected at higher energies ($\Sigma_{Z'}>\Sigma_Z$), the UHECRs composition is expected to be heavier.

\item the surviving nuclei $Z^{'}$ in the energy range between $E_1$
  and $E_2$. Assuming that the source is more abundant in nuclei $Z'$
  than nuclei $Z$, the source should emit a larger amount of nuclei
  $Z'$ in the energy range $E_1$ and $E_2$. So after propagation, the
  number of surviving nuclei in a fixed energy range is $\propto
  f_{Z,\rm surv.}\int k_ZE^{-s}dE \propto f_{Z,\rm
    surv.}M_ZZ^{s-2}$. With the fact that heavier nuclei lose nucleons
  slower than lighter nuclei at the same energy (not the same Lorentz
  factor), we expect the ratio of nuclei $Z'$ and nuclei $Z$ emitted
  by the source in the energy range to be
\begin{equation}
\frac{N_{Z',\rm prop}(E_1;E_2)}{N_{Z,\rm prop}(E_1;E_2)}=\frac{M_{Z'}}{M_Z}\frac{f_{Z',\rm surv.}(E_1;E_2)}{f_{Z,\rm surv.}(E_1;E_2)}\left(\frac{Z'}{Z}\right)^{s-2} > 1
\end{equation}
It does not mean however that these $Z'$ nuclei would contribute to
the anisotropy pattern seen in the range $[E_1,E_2]$, because they
have smaller rigidity than the $Z$ nuclei.

\end{itemize}

If the source does not accelerate nuclei beyond charge $Z$, then the
anisotropy at higher energies is produced by nuclei of charge $Z$. To
derive the corresponding ratio of significances, make the substitution
$M_{Z'}/M_Z\rightarrow 1$, $Z'/Z\rightarrow E_1'/E_1$ in Eq.~(29). Then, with
$(p_2-3)/2\simeq 0.65$, one expects the ratio to increase slightly up
to the energy at which the distance to the source matches the energy
loss distances, then to drop sharply beyond this distance. The
detection of such a feature would provide useful constraints on $Z$
and $D$.

\section{Conclusion}\label{sec:conc}
In this work, we have generalized a test of the chemical composition
of UHECRs, which proposes to use the anisotropy pattern measured as a
function of energy. The basic principle is that if anisotropies are
observed at high energies $E\sim 6\times 10^{19}\,$eV, and if one
assumes that these anisotropies are caused by heavy nuclei of charge
$Z$, then one should observe a strong anisotropy signal at energies
$E/Z$ close to the ankle, due to the proton component
\citep{Lemoine09}. In the present paper, we have accounted for the
production of secondary protons through the photodisintegration
interactions of nuclei. Assuming that no anisotropy signal is detected 
at low energies, we derive an upper bound on the distance to the source.

Our numerical estimates are based on the report of the Pierre Auger
Observatory of an excess in the direction to Cen~A. At present, the
significance of this detection is not well established and one must
await future data to confirm or invalidate it. Nevertheless, the
method presented here remains general and might well be applied to
future more extended datasets. Taking the results of the Pierre Auger
Observatory at face value, we derive a maximal distance to the source
of order $20-30\,$Mpc, $80-100\,$Mpc, $180-200\,$Mpc if the nuclei
responsible for the anisotropies are oxygen, silicon or iron
respectively. The differences between these estimates of the maximal
distance are directly related to the energy loss lengths of these
nuclei at GZK energies. Our results are summarized in
Fig.~\ref{metallicity}, which shows the minimum mass of metals
relatively to hydrogen required in the source, in order to produce a
weaker anisotropy at $E/Z$ than at $E$. At distances exceeding the
above estimates, this amount diverges, meaning that even if the source
does not accelerate any protons, the amount of secondary protons
produced during propagation is sufficient to cause a secondary
anisotropy at $E/Z$ larger than that observed at $E$. At small source
distances, where photodisintegration effects are negligible, one
nevertheless finds a minimum mass $M_Z/M_{\rm H}\,\gtrsim\,1$. When
measured relatively to the solar composition, this indicates that the
metallicity inside the source should exceed $\sim 120Z_\odot$, $\sim
1600Z_\odot$, or $\sim 1100Z_\odot$ if oxygen, silicon or iron nuclei
are responsible for the high energy anisotropy. This result does not
depend on the spectral index, or on the details of the injection
spectrum, as long as the latter is shaped by rigidity. When combined,
these bounds on the distance and metallicity bring in quite stringent
constraints on the source of these particles. Additionally, these
constraints imply that if the heavy nuclei at GZK energies are silicon
or iron, the proton fraction in the all-sky composition at ankle
energies should be less than $\sim50\,$\%, in potential conflict with
measured data.

\acknowledgments We thank the anonymous referee for his/her valuable suggestions. 
This work is supported by the 973 program under grant
2009CB824800, the NSFC under grants 11273016,  10973008, and
11033002, the Excellent Youth Foundation of Jiangsu Province
(BK2012011), and the Fok Ying Tung Education Foundation.

\end{document}